\documentclass[11pt]{article}
\usepackage{amsmath}   
\usepackage{amssymb}
\usepackage{wasysym}
\usepackage{graphicx}
\usepackage[top=1.15in,bottom=1.15in,left=0.9in,right=0.9in]{geometry}
\usepackage[utf8]{inputenc}
\usepackage{cleveref}
\usepackage{authblk}
\crefname{section}{§}{§§}

\newtheorem{lemma}{Lemma}

\begin{document}

\title{Red-Black Trees with Constant Update Time}

\author[1]{Amr Elmasry}
\author[2] {Mostafa Kahla}
\author[2] {Fady Ahdy}
\author[2] {Mahmoud Hashem}
\affil[1]{Department of Computer Engineering and Systems, Alexandria University, Egypt}
\affil[2]{Computer and Communications Engineering Program, Alexandria University, Egypt}
\date{}

\maketitle
	
\begin{abstract}
    We show how a few modifications to the red-black trees allow for $O(1)$ worst-case update time (once the position of the inserted or 
		deleted element is known). The resulting structure is based on relaxing some of the properties of the red-black trees while guaranteeing 
		that the height remains logarithmic with respect to the number of nodes. Compared to the other search trees with constant update time, our 
		tree is the first to provide a tailored deletion procedure without using the global rebuilding technique. In addition, our structure is very simple to 
		implement and allows for a simpler proof of correctness than those alternative trees.
	\end{abstract}
	
	\section{Introduction}
	
    Balanced search trees are among the most common and well-studied data structures used in various
    algorithms. They provide an elegant solution to the dictionary problem, in which one needs to maintain a dynamic set of elements.
    Searching, inserting, and deleting an element is carried out in $O(\log n)$ time, 
		where $n$ is the number of elements in the tree.
    There exists a great variety of balanced search trees in the literature.	
    Examples include height-balanced trees \cite{Knuth1998}, weight-balanced trees \cite{NievergeltR73}, B-trees \cite{Bayer72,BayerM72}, and red-black trees \cite{GuibasS78}.
    Guibas and Sedgewick \cite{GuibasS78} introduced the red-black tree as a fundamental balanced search tree, which was
    derived from the symmetric binary B-tree \cite{Bayer72}. Andersson \cite{Andersson93} simplified the insert and delete operations for red-black trees. Sedgwick \cite{Sedgewick_left-leaningred-black} gave a variation that he calls left-leaning red-black trees. 
		Okasaki \cite{Okasaki99} showed how to make the insert operation purely functional.
		Tarjan \cite{Tarjan83} gave a version of red-black trees that performs a constant number of structural changes (rotations) per operation.
		Still, the worst-case update time for Tarjan's trees is logarithmic.

	Normally, any insertion or deletion in balanced search trees comprises three steps: 1) Querying the position of the key;
    this step takes $O(\log n)$ time.
	2) Performing the actual insertion or deletion, in $O(1)$ time.
	3) Rebalancing the tree; this step always takes $O(\log n)$ time.
    However, there are some applications (see \cite{Mulmuley91} for example) in which we already know the position of the node to be
    inserted or deleted. In this case, the operational complexity depends on the updating (rebalancing) time.
    
		Guibas et al. \cite{GuibasMPR77}, Huddleston and Mehlhorn \cite{HuddlestonM82}, and Overmars \cite{Overmars82} presented methods
			that achieve O(1) amortized update time once the position of the key is known.
    Levcopoulos and Overmars \cite{LevcopoulosO88} presented a balanced search tree achieving 
		$O(1)$ worst-case update time by using a global splitting lemma that is   
		based on a pebble game combined with the bucketing technique of Overmars \cite{Overmars82}. 
		In particular, instead of storing single keys in the leaves of their search tree, 
		each leaf stores a bucket of an ordered list of several keys. 
		The buckets in \cite{LevcopoulosO88} have size $O(\log^2{n})$ each, so they needed a
    two level hierarchy of lists in order to guarantee $O(\log n)$ query time within the buckets. 
		Additionally, deletions relied on global rebuilding \cite{Overmars83}; a very complex process to be executed incrementally.
    Fleischer presented a simpler approach \cite{Fleischer96} to the problem using properties of the (a-b)-trees and
    the bucketing technique, but he did not give a direct deletion technique and also relied on global rebuilding. 
		In addition, his proof of correctness is complicated.
		
		In this paper we give a red-black tree with $O(1)$ worst-case update time per operation.
    Our red-black-tree-based approach is quite similar to Fleischer's implementation in the sense
    of inserting nodes into buckets not in the internal tree and splitting the buckets if they become large. 
		The splitting operation results in adding a new node to the internal tree, and may cause a violation of the original
		red-black tree properties, but we will show that we can fix these violations before the bucket splits again. 
		For deletion in our trees there is no need for global rebuilding like in \cite{Fleischer96,LevcopoulosO88}, 
		and accordingly the structure requires less space and time.
		The paper is organized as follows: Section 2 demonstrates our data structure, Section 3 proves the correctness of our tree, and
    Section 4 is the conclusion.

	\section{The Data Structure }
	A red-black tree, as described in \cite{GuibasS78,Andersson93}, is a binary search tree with one extra bit per node that represents its color (either red or 
	black). Each node has several attributes: key, left-child pointer, right-child pointer, parent pointer, and color bit. If a 
	child of a node does not exist, we make the corresponding child pointer point to a dummy leaf with an arbitrary value.
	The black height of a node $x$ is defined as the maximum number of black nodes on a path from $x$ to any of its descendant leaves.
	
	The following properties must be satisfied for a red-black tree: \\
	\\\textbf{(1)} Every node is either red or black.
	\\\textbf{(2)} The root is black.
	\\\textbf{(3)} Every leaf is black.
	\\\textbf{(4)} If a node is red, then both its children are black.
	\\\textbf{(5)} For each internal node, the black height of its two children is the same.
	\\
	
	In this section we present a slightly modified red-black tree that uses the bucketing technique (each leaf is a node that points to a sorted list of nodes) and must satisfy the following properties: \\
\\\textbf{(1)} Every node is either red or black.
\\\textbf{(2)} The root is black.
\\\textbf{(3)} Every leaf is a bucket and is black.
\\\textbf{(4)} If a node is red and has a red parent, then both its children are black.
\\\textbf{(5)} For each internal node, the difference between the black height of its two children is at most one.
\\

	A major difference between our trees and the vanilla red-black trees is that the leaves of our trees are buckets.
	We have also relaxed property 
	(4) for the vanilla red-black trees and allowed pairs of consecutive red nodes on any path. 
	We consider this red pair as a {\it double-red violation} that needs to be incrementally fixed within the upcoming operations.
	The last difference is that property (5) in our trees states that for each node the black height of its two children differs by at most one, 
	while in the vanilla trees property (5) states that for each node both its children must have the same black height. 
	When a black node has a black height that is one less than that of its sibling, to compensate for this difference we deal with its color as 
	doubly black. This {\it doubly-black violation} is to be fixed incrementally within the upcoming operations. 
	 The following lemma bounds the height of our trees.
	
	\begin{lemma}
	The height of a tree that has $n$ internal nodes is at most $4.32 \log_2(n+2)$.
	\end{lemma}	
	\noindent{Proof.}
	By considering property (5), the black height of any node resembles the height of an AVL tree, which is at most $1.44 \log_2(n+2)$. 
	Also, by considering property (4), the number of black nodes on any path is at least one third of the total number of nodes, and thus the 
	height of the tree is at most $3 \times 1.44 \log_2(n+2) = 4.32 \log_2(n+2)$.
	\hfill $\Box$ \\
	
Let $H = \lceil 4.32 \log_2(n+2) \rceil$ denote an upper bound for the height of a tree of $n$ internal nodes, in accordance with the above lemma. We maintain the following invariant on the size of our buckets. 
\\
\\\textbf{(6)} The number of nodes $L_b$ in every bucket $b$ must be in the range $0.5H \leq L_b \leq 2H$.
\\
	
	When querying for an element, we follow the search path from the root until reaching a bucket that we scan looking for the element. 
	For insertions and deletion, we assume that the location of the key to be inserted or deleted is known. 
	A fixing pointer $\gamma_b$ is associated with every bucket $b$ and points to a node on $b$'s search path. 
	A fix-up procedure is called within every updating operation and utilizes the fixing pointer $\gamma_b$ of the bucket $b$ where the update 
	took place. Let $V$ be the node pointed to by $\gamma_b$ before the fix-up procedure. 
	The following fix-up procedure is executed if $V$ is not the root.
	
	\begin{enumerate}
	\item If $V$ is red with a red parent, fix $V$ using the {\it double-red fix-up} procedure (\cref{sec:double-red})
	\item If $V$ is doubly-black, fix $V$ using the {\it doubly-black fix-up} procedure (\cref{doubly-black})
	\item Advance $\gamma_b$ to point to $V$'s parent.
	\end{enumerate}
	
	Note that the double-red and the doubly-black fix-up procedures are guaranteed to either eliminate the violation at node $V$ or propagate it to $V$'s parent or $V$'s grandparent. In accordance, the violation will be fixed in at most $H + O(1)$ steps. 
	
	\subsection{Insertion}
	\label{sec:insertion}
\begin{figure}[ht!]
		\centering
		 \includegraphics[width=0.4\linewidth]{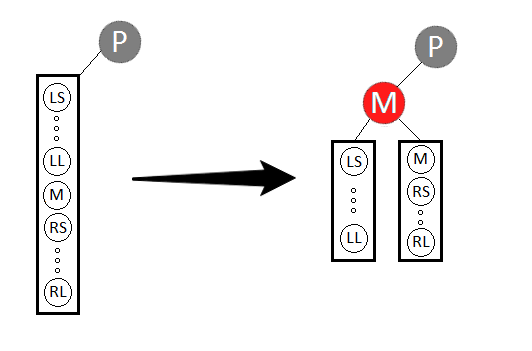}
		\caption{Splitting a bucket. The middle element M is now added to the tree as a red node. As in all other figures, note that the gray nodes could be possibly black or red.}
		\label{fig:split}
	\end{figure}

	Our insertion algorithm is similar to Fliecher's \cite{Fleischer96}. One difference is that we split a bucket when the number of its nodes is about to 
	overstep that of property (6), and not when its fixing pointer reaches the root. 
	We perform the insertion according to the following algorithm:
	
	\begin{enumerate}
	\item Insert the new node at the designated location of the designated bucket $b$.
	\item Call the fix-up procedure for $b$.
	\item If the bucket's size $L_b$ exceeds $2H-10$:
	  \begin{enumerate}
	    \item If $\gamma_b$ has not reached the root yet: Repeatedly call the fix-up procedure for $b$ until $\gamma_b$ reaches the root. 
		  \item Split the bucket into two buckets as shown in Figure \ref{fig:split}. Add their new parent as an internal node in place of the old bucket, color it red, and make the fixing pointer of both new buckets point to this node. Call the fix-up procedure for one of the two buckets (to instantaneously dismiss a possible double-red violation from above the buckets).
	   \end{enumerate}
	\end{enumerate}	 
	
	Since the bucket's size after splitting is at most $H$ and after merging is less than $H+6$ (see \cref{sec:deletion}) 
	and the bucket is split when its size exceeds $2H-10$, 
	it then takes $H-O(1)$ insertions in this bucket before it is split (as long as $H$ has not changed).
	As we typically call the fix-up procedure once per insertion,
	the fixing pointer of the bucket would be at a constant distance from the root when the size of the bucket reaches the threshold.
	Alternatively, as the value of $H$ infrequently decreases, 
	it is possible that the fixing pointer would need extra fix-ups to reach the root. 
	But, as we illustrate in \cref{sec:global}, 
	the number of the extra fix-ups needed until the fixing pointer reaches the root is also a constant. 
	
	\subsection{Deletion}
	\label{sec:deletion}
\begin{figure}[ht!]
    \centering
    \includegraphics[width=0.4\linewidth]{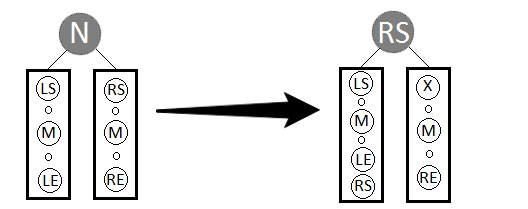}
    \caption{The left bucket borrows a node from the right bucket by adding the right bucket's first node RS to the end of the left bucket. The parent will have a new value equals RS. Node X, which follows RS, is now the first node in the right bucket.}
    \label{fig:borrow-1}
\end{figure}

\begin{figure}[ht!]
    \centering
    \includegraphics[width=0.4\linewidth]{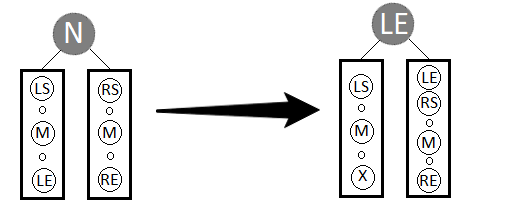}
    \caption{The right bucket borrows a node from the left bucket by adding the left bucket's last node LE to the start of the right bucket.
		The parent will have a new value equals LE. Node X, which precedes LE, is now the last node in the left bucket.}
    \label{fig:borrow-2}
\end{figure}

\begin{figure}[ht!]
		\centering
		\includegraphics[width=0.4\linewidth]{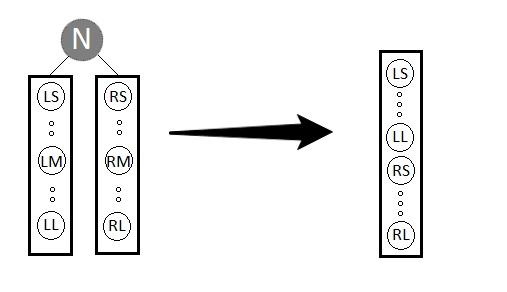}
		\caption{Merging two buckets. The elements of the right bucket are appended after the tail of the left bucket, and the new merged bucket's parent will be N's parent.}
		\label{fig:merge}
	\end{figure}

	In a binary search tree, the deletion of an element in an internal node that has two children usually requires replacing the element with its 
	successor. To locate the successor in $O(1)$ worst-case time, we can augment each node with a pointer to its successor bucket and when we want to delete this node we replace it with the smallest node in its successor bucket. Deleting any node is thus equivalent to removing a node from a bucket. For simplicity, for the rest of the paper, we assume that all data items are stored in the buckets and that the internal nodes only have key indexing information.
	We perform the deletion according to the following algorithm:
	
	\begin{enumerate}
	\item Remove the desired node from the bucket $b$.
	\item Call the fix-up procedure for $b$ twice.
	\item If the bucket's size $L_b$ is less than $0.5H+3$:
	 \begin{enumerate}
	  \item If $\gamma_b$ has not reached the root yet: Repeatedly call the fix-up procedure for $b$ until $\gamma_b$ reaches the root. 
	  \item If $b$'s sibling bucket $b'$ has $L_{b'} > 0.5H+3$: Borrow a node from $b'$ as shown in Figures \ref{fig:borrow-1} and \ref{fig:borrow-2}. Call the fix-up procedure for $b'$ twice. 
	  \item If $b$'s sibling bucket $b'$ has $L_{b'} \leq 0.5H+3$: Merge $b$ and $b'$ in bucket $b''$ as shown in Figure \ref{fig:merge}. 
		If the deleted common parent was black, mark $b''$ as doubly-black. Make $\gamma_{b''}$ point to $b''$.		
	\end{enumerate}
\end{enumerate}	

Since the bucket's size after merging is at least $H$ and after splitting is at least $H-5$ (see \cref{sec:insertion})
and the bucket will be merged when its size is less than $0.5H+3$, 
it then takes $0.5H - O(1)$ deletions in this bucket before it is merged (as long as $H$ has not changed). 
So, we move the bucket's fixing pointer twice up per deletion
for the bucket's pointer to be at a constant distance from the root by then.
Alternatively, as the value of $H$ infrequently increases, it is possible that the fixing pointer would need extra fix-ups to reach the root. 
But, as we illustrate in \cref{sec:global}, 
the number of the extra fix-ups needed until the fixing pointer reaches the root is also a constant. 

\begin{figure}[ht!]
		\centering
		\includegraphics[width=0.4\linewidth]{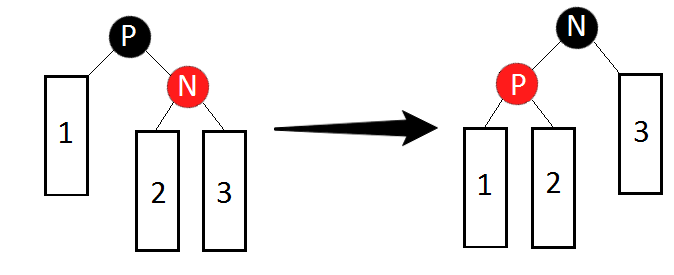}
		\caption{Bucket 1 has a sibling node that is not a bucket. In order to merge with or borrow from bucket 2, we do a left rotation for P.
		(Alternatively, if bucket 1 was the right child, we rotate P right.)}
		\label{fig:bucketSibling}
	\end{figure}

Note that it is not always the case for a bucket's sibling to be a bucket. However, since the bucket's black height is 1 then its sibling must have black height 1. The reason is that this sibling can't have black height 0 by property (3) (as buckets have height 1), and it can't have black height 2 also as this means that the bucket itself would be doubly black and this is impossible. Also, step 3(b) in the insertion procedure guarantees that we can't have red parent and grandparent for a bucket. It follows that Figure \ref{fig:bucketSibling} shows the only way for a bucket to have a node sibling and how with a single rotation we are guaranteed to have a sibling bucket. 
	
\subsection{Global fixing}
\label{sec:global}
Since $H$ is a function of the number of internal tree nodes $n$ that continuously changes, this may cause some bucket sizes to disobey the thresholds not as a result of an insertion or deletion but because $H$ changes. 
This problem did not emerge in  \cite{Fleischer96} and \cite{LevcopoulosO88} as their bucket sizes are only bounded as a function of $n$ from above. Since their trees, lacking a support for real time deletion, were only growing in size, a violation of the bound for a bucket takes place only by insertions in the bucket. However, in our tree the bucket sizes are bounded from below and from above. Since we allow both insertions and deletions, $n$ can increase or decrease, and as a result this boundary may be violated for some buckets just because $H$ changes even if no operations take place at those buckets.
We compute $H$ as $\lceil 4.32\log_2{(n+2)} \rceil$. 
The value of $H$ increases by at most five when $n$ doubles and decreases by at most five when $n$ halves, 
and the value of $0.5H$ increases or decreases by at most three.
If the above issue is left unresolved, property (6) may be dramatically violated as $H$ repeatedly increases or decreases. 
	
Another problem we might face as $H$ changes is that some bucket sizes could reach the threshold while their fix-up pointers are not yet at the root. When $H$ increases, the size of the buckets relative to $0.5H$ decreases by three. 
In this case, for those buckets, we are short by the effect of the fix-ups of three deletion operations and need to make up for six additional fix-ups (two per missing deletion). Also, since $H$ increases by five, it is possible that up to five levels are added to the path between their fixing pointers and the root. Therefore, each of those buckets may need up to a total of eleven fix-ups for their pointers to reach the root. Alternatively, when $H$ decreases, the size of the buckets relative to $2H$ increases by ten. 
In this case, for those buckets, we are short by the effect of ten insertion operations and need to make up for ten additional fix-ups. 
With time, the number of additional fix-ups needed to reach the root will stack up with each increase or decrease in $H$. 

To resolve these issues, we scan over the buckets while performing the insertions and deletions operations. 
Within this scan, we call the fix-up procedure eleven times for each of two buckets per operation. 
In addition, we fix by splitting or borrowing/merging the buckets that have sizes more than $2H-10$ or less than $0.5H+3$, respectively.
Since we scan all the buckets before $n$ doubles or halves, all the bucket sizes will always be at least $0.5H$ and at most $2H$
and all the shortage in the fix-ups will be compensated for. 
Eventually, we prove in the Appendix that three fix-ups per operation are enough. 
Actually, as $n$ increases only by bucket splits and decreases only by bucket merges, one can show that $n$ quadruples or is divided by four in $\Omega(n \cdot H)$ operations. It follows that we just need a slower speed (the number of extra fix-ups could be divided by $H$) within the global scan.

One last piece of non-synchronization might happen after $H$ changes and the size of a bucket reaches the threshold of splitting or merging before doing the aforementioned global fix-ups. To resolve this issue, within the insertion (\cref{sec:insertion}) and deletion (\cref{sec:deletion}) operations, we perform at most eleven fix-ups until the bucket's pointer is at the root once its size violates the threshold.
	
\subsection{Operations inside the buckets}

To perform the merging and splitting of the buckets in constant worst-case time, we maintain pointers to the first, middle and last nodes of each bucket. Updating the three pointers when two buckets are merged is straightforward. It is also straightforward to update the middle pointer after insertions and deletions, by moving the pointer one step once the count of nodes on one side of the pointer exceeds the other by two. When a bucket splits, both the resulting buckets will not have the middle pointers correctly set. We adjust the middle pointers incrementally, by initializing a tentative pointer to the beginning of each of the two bucket and moving the pointer forward with every updating operation within the bucket while keeping track of the count of nodes on both sides of the pointer. Since the number of operations that will be performed until the next split is at least as many as the required pointer displacement steps, it is guaranteed that the middle pointer will be correctly set before then.
		
		When we insert or delete a node we need to access the fixing pointer of the bucket in which this node resides in order to perform the fix-up procedure. For that, each node needs to refer to the fixing pointer of its bucket. But if we merge or split a bucket, these references need to be changed for all nodes. To resolve this issue, we keep two identical copies of the fixing pointers with each bucket, the left copy is supposed to be referenced by the nodes that precede the bucket's middle pointer and the right copy is supposed to be referenced by the nodes that follow the middle pointer. When a node is inserted it is made to refer to the corresponding copy of the fixing pointer.
Accompanying each updating operation within the bucket, when the middle pointer is moved, we redirect the affected nodes to refer to the corresponding copy of the fixing pointer. When a bucket is split, the left copy is kept with the left bucket and the right copy is kept with the right bucket. All the nodes of each of the two split buckets will be in accordance referring to the corresponding copy. A new left copy of the fixing pointer is created for each of the two buckets, and the nodes preceding the middle pointer are made to refer to the new copies one by one. When two buckets are merged, we will subsequently have four copies of the fixing pointer instead of two. To return back to the two-copies situation, with each updating operation, we incrementally redirect the references of two of the nodes of the left side to point to one of the first two copies and two more of the right side to refer to one of the other two copies. Once all the nodes of each side point to one copy, we discard the other copy. At the end, we are left with two copies, and this happens before the next merge or split operation is possibly due.

\subsection{Double-red fix-up procedure}
\label{sec:double-red}
The double-red fix-up procedure in the vanilla red-black tree aims to handle a red node that has a red parent as this violates property \textbf{(4)}. The procedure \cite{Cormen98,Tarjan83} resolves the violation either by $O(1)$ rotations if possible, or by a color flip for the parent to become black, so property \textbf{(4)} is no longer violated for this node but the node's grandparent becomes red and may violate the property again. This way, another fix is to be performed and possibly repeated, making the procedure run in $O(H)$ time. In our tree, we distribute the work over the next $O(H$) operations so that each operation takes $O(1)$ update time. Later in \cref{sec:correctness} we prove that this lazy re-balancing will not invalidate the invariants of our tree. Our double-red fix-up procedure is almost identical to that of the vanilla red-black tree. For completeness, we give the procedure of the vanilla tree first then explain our modifications.

\begin{figure}[ht!]
	\centering
	\includegraphics[width=0.4\linewidth]{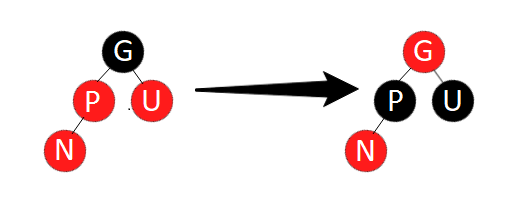}
	\caption{Double-red fix-up: Case 1}
	\label{fig:insertion-1}
\end{figure}

\begin{figure}[ht!]
	\centering
	\includegraphics[width=0.5\linewidth]{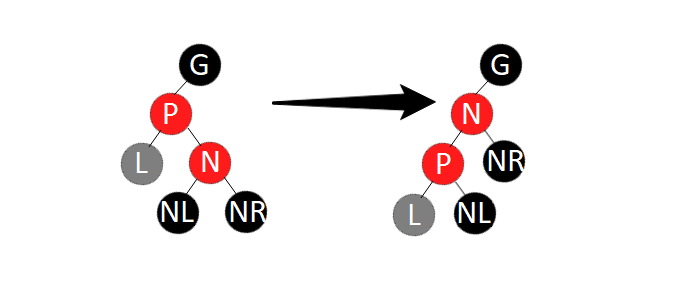}
	\caption{Double-red fix-up: Case 2}
	\label{fig:insertion-2}
\end{figure}

\begin{figure}[ht!]
	\centering
		\includegraphics[width=0.4\linewidth]{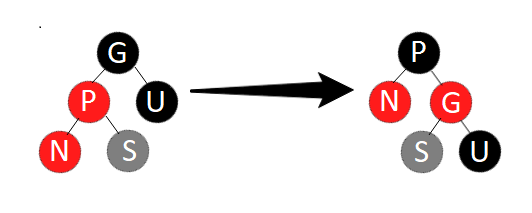}
	\caption{Double-red fix-up Case 3}
	\label{fig:insertion-3-a}
\end{figure}

We call this procedure for a red node (N) that has a red parent (P) and parent's sibling (U) and grandparent (G). In each iteration, this procedure is guaranteed to resolve the violation or propagate it to the node's grandparent. 
We only show cases where node P is the left child of G, the other cases are symmetric. 
The following is the vanilla red-black tree double-red fix-up:

\begin{description}
\item{Case 1}: 
If U is red, then color both P and U black and color G red. If G was the root, it stays black. See Figure \ref{fig:insertion-1}.
\item{Case 2}: 
If U is black and N is the right child of P, then perform a left rotation at P. Now P is red and its parent N is red and P is the left child of N. We then proceed with Case 3. See Figure \ref{fig:insertion-2}.
\item{Case 3}: 
If U is black and N is the left child of P, then perform a right rotation on G, switch the colors of P and G and terminate. See Figure \ref{fig:insertion-3-a}.

\end{description}

\begin{figure}[ht!]
	\centering
	\includegraphics[width=0.45\linewidth]{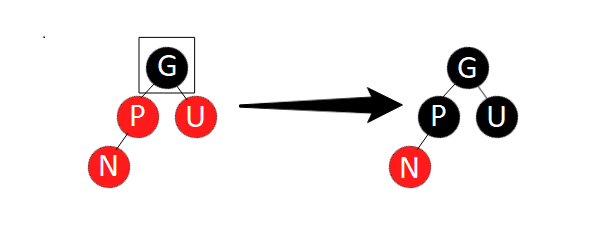}
	\caption{Double-red fix-up: Case 1.1}
	\label{fig:insertion-1.5}
\end{figure}
\begin{figure}[ht!]
	\centering
	      \includegraphics[width=0.4\linewidth]{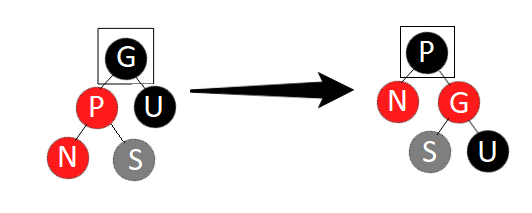}
	    \caption{Double-red fix-up: Case 3.1}
	\label{fig:insertion-3-b}
	\end{figure}

It is clear that all cases do not invalidate the properties and do not create new doubly-black nodes or double-red nodes except for Case 1, which will create a double-red pair if G's parent was red.

Our modification of this procedure is to do one iteration and then terminate allowing double-red nodes to exist. 
Since we allow doubly-black nodes in our tree, we need to make other modifications.
If G is doubly-black in Case 1.1, we will not color it red. Instead, we leave G black by removing the doubly-black flag and terminate. This is because both its children are now colored black and its black height has increased by one, and so it is no longer doubly-black. See Figure \ref{fig:insertion-1.5}. Also, we added new case 3.1 in which
if G is doubly-black then we perform a right rotation on G, color P doubly-black and G red and terminate. See Figure \ref{fig:insertion-3-b}.

\subsection{Doubly-black fix-up procedure}
\label{doubly-black}

The doubly-black fix-up procedure in the vanilla red-black tree is executed when the deleted node is replaced by its black successor. This causes the path that previously contained this successor to have one fewer black node, violating property \textbf{(5)}.  The procedure \cite{Cormen98,Tarjan83} resolves the violation either by $O(1)$ rotations if possible, or propagates the doubly-black violation to the node's parent by color flips. This way, another fix is to be performed and possibly repeated, making the procedure run in $O(H)$ time. 
In our case, a bucket is colored doubly-black following a merge operation if the parent of the two merged buckets was black.
Similar to the double-red fix-up, we distribute the work over the next $O(H)$ operations. 
Later in \cref{sec:correctness} we prove that this lazy re-balancing will not invalidate the invariants of our tree.
We give the procedure of the vanilla tree first then explain our modifications.

\begin{figure}[ht!]
	\centering
	\includegraphics[width=0.35\linewidth]{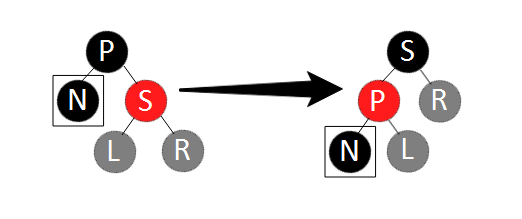}
	\caption{Doubly-black fix-up: Case 1}
	\label{fig:deletion-1}	
\end{figure}

\begin{figure}[ht!]
	\centering
	\includegraphics[width=0.42\linewidth]{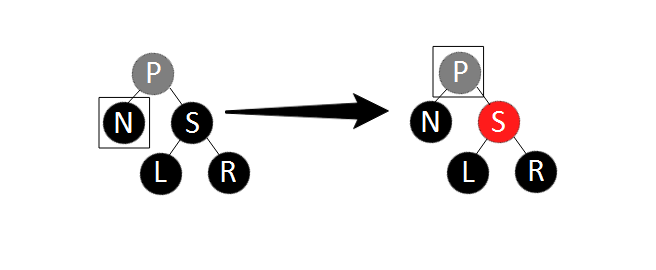}
	\caption{Doubly-black fix-up: Case 2}
	\label{fig:deletion-2}
\end{figure}

\begin{figure}[ht!]
	\centering
	\includegraphics[width=0.48\linewidth]{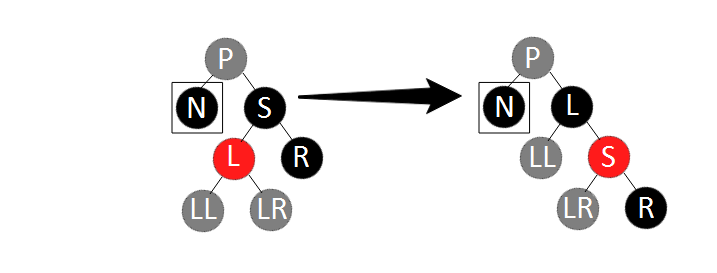}
	\caption{Doubly-black fix-up: Case 3}
	\label{fig:deletion-3}
\end{figure}

\begin{figure}[ht!]
	\centering
	\includegraphics[width=0.4\linewidth]{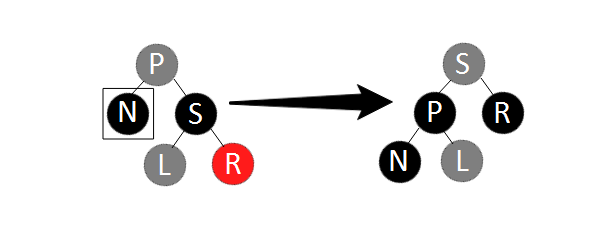}
	\caption{Doubly-black fix-up: Case 4}
	\label{fig:deletion-4}
\end{figure}

We call this procedure for a doubly-black node (N) that has a parent (P) and sibling (S). 
In each iteration, this procedure is guaranteed to either eliminate the doubly-black violation or to propagate it upward. 
We only show cases where node N is the left child of P, the other cases are symmetric.
The following is the vanilla red-black tree doubly-black fix-up: 

\begin{description}
\item{Case 1}:
If S is red and P is black, then perform a left rotation on P, color S black and P red, and proceed to the next cases. See Figure \ref{fig:deletion-1}.

\item{Case 2}:
If S is black and both its children are black, then color S red:

\begin{description}
\item{(a)}
If P is red, color P black and terminate. 
\item{(b)}
If P is black and is not the root, set its doubly-black flag.
See Figure \ref{fig:deletion-2}.
\end{description}

\item{Case 3}:
If S is black, the right child of S is black and the left child is red, then perform a right rotation on S, exchange its color with its new parent and proceed to Case 4. See Figure \ref{fig:deletion-3}.

\item{Case 4}:
If S is black and its right child is red, perform a left rotation on P, exchange the color of P and S, color the right child of S black, reset N's doubly-black flag and terminate. See Figure \ref{fig:deletion-4}. 

\end{description}


Our modification of this procedure is to do one iteration and then terminate allowing doubly-black nodes to reside.
Since we allow double-red pairs and doubly-black nodes to exist in our tree, the following modifications need to be made.
First, Case 1 would possibly be repeated twice since the new sibling of N after the rotation may also be red.
In addition, we need to add two new cases to Case 1.

\begin{figure}[ht!]
	\centering
		\includegraphics[width=0.4\linewidth]{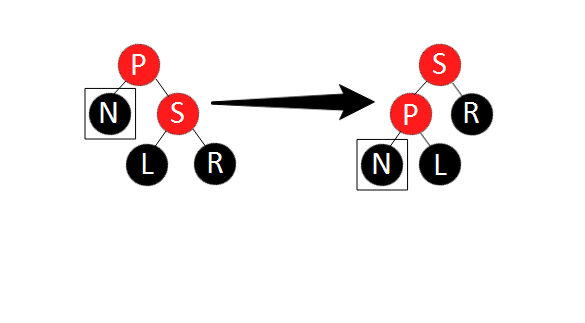}
	\caption{Doubly-black fix-up: Case 1.1}
	\label{fig:deletion-1.1}
\end{figure}

\begin{figure}[ht!]
	\centering
		\includegraphics[width=0.4\linewidth]{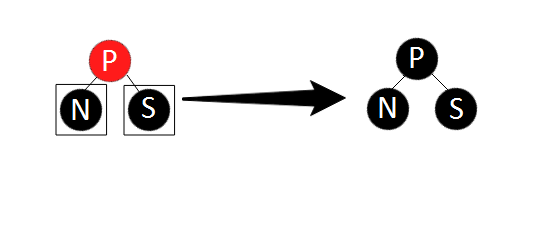}
	\caption{Doubly-black fix-up: Case 1.2(a)}
	\label{fig:deletion-1.2-a}
\end{figure}

\begin{figure}[ht!]
	\centering
		\includegraphics[width=0.4\linewidth]{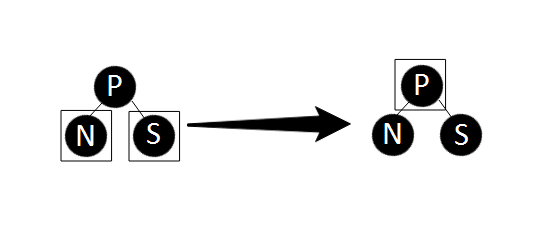}
	\caption{Doubly-black fix-up: Case 1.2(b)}
	\label{fig:deletion-1.2-b}
\end{figure}

\begin{description}
\item{Case 1.1}:
If S is red and P is also red, then perform a left rotation on P but without recoloring and proceed to the next cases. See Figure \ref{fig:deletion-1.1}.
\item{Case 1.2}: 
If S is doubly-black, then remove the doubly-black flag from S and N:
\begin{description}
\item{(a)}
If P is red, color P black and terminate. See Figure \ref{fig:deletion-1.2-a}.
\item{(b)}
If P is black and is not the root, set its doubly-black flag. See Figure \ref{fig:deletion-1.2-b}.
\end{description}
\end{description}


The above procedure will either eliminate the doubly-black violation (Case 4) or propagate it upward to its parent (Cases 1.2(b) and 2). 
Additionally, no new double-red pairs or doubly-black nodes are created.
We prove in \cref{sec:correctness} that both the double-red and doubly-black fix-up procedures will not result in invalidating the invariants of our red-black trees.

\section{Validity of the Tree}
\label{sec:correctness}
In this section we prove that the properties of our trees hold. 
Since we allow double-red and doubly-black violations, and as we do lazy rebalancing, 
we need to demonstrate that these limited violations would not extend forming more drastic violations. 
We need to show that if a node is a black node that has a red parent and a red grandparent, it is impossible for any of its children to propagate a red color to it (by double-red fix-up Case 1), as this will violate property \textbf{(4)}. 
We also need to show that for any node that is doubly black none of its children will propagate another doubly-black violation to it (by doubly-black fix-up Cases 1.2(b) or 2), as this will violate property \textbf{(5)}.
	
We define a bucket to be proper for a node N if its fixing pointer points to one of N's descendants or N itself, otherwise the bucket is non-proper for N. A proper bucket will not generate any new violation (double-red nodes by splitting or doubly-black node by merging) until its pointer passes through N up to the root, fixing the violation at N. 
We define the termination case for a doubly-black violation to be a red node and for a double-red violation to be a couple of consecutive black nodes or a doubly-black node. 
Now our proving pattern for properties \textbf{(4)} and \textbf{(5)} are analogous. 	
We will prove that for any node forming a violation there exists a termination case on each path between this node and a descendant violation of the same type and each path to a descendant non-proper bucket. 
The key point is that no violation can propagate upward to this violation since there is a termination case in between. 
It is straightforward to verify that these termination cases indeed prevent the violations from joining.
Also if a bucket is doubly-black, it cannot be doubly-black again before it is fixed.

\begin{lemma}
For any path from a doubly-black node to a descendant doubly-black node or to a descendant non-proper bucket there exists a termination case for the doubly-black violation, which is a red node.
\end{lemma}

\noindent{Proof.}
We prove our claim by induction on the doubly-black nodes regarding their heights. 
For the base case, the claim holds trivially for buckets. 
As the fixing pointer of a doubly-black bucket must be pointing to itself, it is a proper bucket for others.
When this doubly-black violation propagates upward: i) there is no need for a termination case between the violation and the proper bucket, and ii) any other bucket that was proper for this violation remains proper.
Just before a bucket becomes doubly black its fixing pointer must have been pointing to the root, and the bucket was non-proper for its ancestor nodes. Hence, by induction, there is a red node on the path between this bucket and its first ancestor doubly-black node, if any exists. This red node becomes a termination case between the now doubly-black bucket and the first ancestor doubly-black node.

For the rest of the inductive step, let P be an internal node in the tree. From the doubly-black fix-up procedure, the only way for P to become doubly black is either by Case 1.2(b) or by Case 2. 
In Case 2 (see Figure \ref{fig:deletion-2}), P has two children N and S where N is doubly black. If N is a bucket, then it is a proper bucket. If N is an internal node, then by induction the claim holds for P with any descendant of N. Since S becomes red forming a termination case, the claim holds for P with any descendant of S. Thus, the inductive step follows.
In Case 1.2(b) (see Figure \ref{fig:deletion-1.2-b}), P has two doubly-black children N and S. If any of them is a bucket, then it is a proper bucket. If any of N or S is an internal node, then by induction the claim holds for it and hence for P that becomes doubly black. So, there exists a red node between P and any descendant doubly-black node or non-proper bucket. 
	
Since some double-red or doubly-black fix-up cases can cause color flips or a rotation of a red node outside a path, what is left to show is that none of these cases will affect our claim. In the doubly-black fix-up Cases 1,3,4 (see Figures \ref{fig:deletion-1}, \ref{fig:deletion-3}, \ref{fig:deletion-4}), we do a rotation resulting in removing a red node from a certain path, but this is not a problem since node N in those cases was doubly black, thus between it and any ancestor doubly-black node there exists a red node. Hence, the red node that was removed from the path by the rotation will not invalidate the claim. 
In the double-red fix-up Case 1.1 (see Figure \ref{fig:insertion-1.5}), we propagate a red color to a doubly-black node. 
In the doubly-black fix-up Case 1.2(a) (see Figure \ref{fig:deletion-1.2-a}), we propagate a doubly-black violation to a red node. 
As a result, in both cases, both the red color and the doubly-black violation are eliminated. 
And this is not a problem, since any violation (doubly-black) node has a termination case (red node) between itself and any ancestor violation, and so if both the violation and the termination case are removed the claim still holds.
\hfill $\Box$ 

\begin{lemma}
For any path from a double-red violation to a descendant double-red violation or to a descendant non-proper bucket there exists a termination case for the double-red violation, which is either two consecutive black nodes or a doubly-black node. 
\end{lemma}
	
\noindent{Proof.}
We again prove our claim by induction. The base case is established when a bucket splits, and hence the fixing pointers of both the resulting buckets point to their new common red parent. If this parent has in turn a red parent, a double-red violation is created. The two buckets are then proper for the generated double-red violation.
When this double-red violation propagates upward: i) there is no need for a termination case between the violation and the proper buckets, and ii) any other bucket that was proper for this violation remains proper.
Just before a bucket splits generating a double-red violation, its fixing pointer must have been pointing to the root and the bucket was non-proper for its ancestor nodes. Hence, by induction, there is either two consecutive black nodes or a doubly-black node between this bucket and its first ancestor double-red violation, if any exists. This becomes a termination case between the new double-red violation and its first ancestor double-red nodes.

For the inductive step, the only way for a double-red violation to propagate is by double-red fix-up Case 1. 
In this case (see Figure \ref{fig:insertion-1}), node G is black and node N is red and has a red parent P, i.e., N and P form a double-red violation. After the fix-up, G becomes red and P becomes black. If the parent of G is red, the violation at N propagates upward to G.
By induction, the claim holds for N and hence for G with any descendant of N. 
Node U is the other child of G and is red.
If a child of U is red, it formed a double-red violation with U. Hence, the claim follows by induction for G with the descendants of this child of U. If a child of U is black, as U becomes black after the fix-up, we get a termination case of two consecutive black nodes formed by U and this child. Thus, the inductive step follows for G. 
	
Since some cases can cause color flips or node rotations outside a path, we show next that none of these cases will affect our claim.
In the double-red fix-up Case 2 (see Figure \ref{fig:insertion-2}), a double-red violation changes its path. 
However, this case is followed by the execution of Case 3 that remedies the situation by breaking the violation.
In the double-red fix-up Cases 3 and 3.1 (see Figure \ref{fig:insertion-3-a}, \ref{fig:insertion-3-b}), nodes G and U were black, and G is turned red. But since nodes N and P were red forming a double-red violation, there exists another ancestor termination case below any ancestor violation.
In the double-red fix-up Case 1.1 (see Figure \ref{fig:insertion-1.5}), node G was doubly-black and after the fix-up G and both its children are colored black. If G was a doubly-black termination case between two double-red violations, G and any of its children still form a termination case between the two violations. 
In the doubly-black fix-up Case 2 (see Figure \ref{fig:deletion-2}), node S is colored red breaking a two consecutive black pair that forms a possible termination case. If P or its parent were black before the fix-up, a new termination case now exists at P. Alternatively, if both P and its parent were red, a double-red violation is resolved by the fix-up and this permits breaking the termination case.  
Also for the doubly-black fix-up Case 3 (see Figure \ref{fig:deletion-3}), a black node is colored red breaking a two consecutive black pair that forms a possible termination case. However, this case is followed by Case 4 that remedies the situation by creating a new black pair.
In the doubly-black fix-up Cases 4 and 1.2(a) (see Figures \ref{fig:deletion-4}, \ref{fig:deletion-1.2-a}), a doubly-black node is turned black and its parent becomes black. As above, a possible doubly-black termination case is replaced by two consecutive black nodes forming a replacement termination case. 
In all cases, the claim still holds.
\hfill $\Box$ 



	
\section{Conclusion}
We have demonstrated a simple search tree, a variant of the red-black trees, that supports queries in $O(\log n)$ worst-case time 
and updates in $O(1)$ worst-case time once the location of the item is located. 
We have implemented our structure and verified its correctness and practical efficiency.
We expect our ideas can be carried to other forms of search trees like 
AA-trees and left-leaning red-black trees.

\bibliographystyle{plain}
\bibliography{references}

\appendix
\section{Appendix}

We prove next that instead of doing eleven fix-ups for each of two buckets after each operation in the global fixing of \cref{sec:global}, three fix-ups are sufficient. 

Just before $H$ increases each bucket has size at most $2H-10$. This means that just after $H$ increases each of these buckets will have size at most $2H-20$ (when $H$ increases the size of the buckets relative to $2H$ decreases by ten). Therefore, for $H$ to increase again (i.e., for $n$ to double), $n$ splits must take place. With splitting threshold as $2H-10$, each split requires at least eleven insertions in this bucket (from size $2H-20$ till it exceeds $2H-10$). Therefore, in total more than $11 n$ operations must take place before $H$ increases again. Thus, if we do one fix-up in the global scanning after each operation, we will have more than eleven fix-ups for each bucket. 

Alternatively, before $H$ decreases each bucket has size at least $0.5H+3$. Just after $H$ decreases each bucket will have size at least $0.5H+6$ (when $H$ decreases the size of the buckets relative to $0.5H$ increases by three). Merging happens when a bucket has size $0.5H+2$ and its sibling has size $0.5H+3$. Hence, for a merge to take place after $H$ decreases, seven deletions must occur in the two sibling buckets so that the two buckets reach the threshold for merging. For $H$ to decrease again (i.e., for n to be halved), at least $n/2$ merges must take place and so at least $7n/2$ operations must occur. If we perform three fix-ups in the global scanning after each operation, we will have an overall more than $10n$ fix-ups, and hence more than ten fix-ups for each bucket.

\end{document}